%Paper: hep-ph/9301286
%From: ZHANGX%umdhep.BITNET@VTVM2.CC.VT.EDU
%Date: Wed, 27 Jan 93 11:21 EST
%Date (revised): Sun, 31 Jan 93 16:35 EST

\font\titlefont = cmr10 scaled \magstep2
\magnification=\magstep1
\vsize=20truecm
\voffset=1.75truecm
\hsize=14truecm
\hoffset=1.75truecm
\baselineskip=20pt

\settabs 18 \columns

\def\b{\bigskip}
\def\bb{\bigskip\bigskip}

\def\ce{\centerline}

\def\no{\noindent}

%$$\eqalign{
% put in lines of equations here, each ending in \cr
%}$$

%$$\eqalign{
%put in equations here ending each line with \cr
%} \eqno (1)$$
%the above will put the one between the two lines of equations and set it
%off to the right

% END BEGINNING FORMATS
% BEGIN HEADER

 \rightline{ UMDHEP 93-126}
\rightline{ Jan. 1993}
\bb

\b
\ce{\titlefont{Restrictions on B-L Symmetry Breaking}}
\ce{\titlefont{ Implied by a Fourth Generation Neutrino
    \footnote\dag{\rm{ Work supported by a grant from the National
          Science Foundation}} }}
\bb
\ce{\bf{R.N. Mohapatra and X. Zhang
   }}

\ce{\it{ Department of Physics and Astronomy}}
\ce{\it{University of Maryland}}
\ce{\it{ College Park, MD 20742 }}
\b
\ce{\bf Abstract}

\no We point out that if the fourth generation neutrino is a Majorana
fermion,
then the astrophysical
constraints coupled with the precision measurements of the
Z-width at LEP require that
 the corresponding B-L symmetry
is unlikely to be a spontaneously broken
global symmetry. If B-L is chosen to be a local symmetry, its breaking scale
should be less than a few TeV.

 \filbreak
\no {\bf 1.} The recent precision measurements of Z-width[1] at LEP and SLC
have shown
that if there exists a fourth generation of leptonic doublet, the masses of
both the charged lepton ($L^-$) and
neutrino ($\nu_4$) must be heavier than about 45 GeV. They do not alter
the
standard big bang nucleosynthesis
scenario[2]
because due to
their large mass, they
annihilate away much earlier than the
epoch of nucleosynthesis.
It is therefore conceivable that a heavy fourth generation
of quarks and leptons
exists and remains to be discovered.
 It is then important to theoretically probe
the constraints on the properties of these particles. In this brief note,
 we focus on the nature of the
fourth generation neutrino, $\nu_4$.

First it is important to realize that unlike the case for
the first three neutrino species $\nu_e,~~ \nu_\mu$ and
$\nu_\tau$, the above mentioned mass constraint
implies that, if there is a fourth generation
neutrino, it $\underline{must}$ be
accompanied by
a right-handed neutrino
(to be denoted by $\nu_{4 R}$)
 to give it a mass. Note that, if a Higgs triplet with
non-zero vev (vacuum expectation value) were to be
  introduced to give a Majorana mass to the
$\nu_{4 L}$, (in which case
$\nu_{4 R}$ would not be needed,) the vev
of the triplet must be so large that it would
contradict bounds on it implied by the $\rho$ parameter measurement.

The next question that arises is, whether $\nu_4$ is a Majorana or
Dirac fermion. If it is a Dirac fermion, the only constraint on its mass
is that which followes from the precision measurement of the
$\rho$ parameter $\it {i.e.}$
$${
| m_{\nu_4}^2 - m_{L_4}^2 | \leq {( 152~ GeV)}^2  }
\eqno(1)$$

\no On the other hand if it is of Majorana type, the
constraints depend on how its Majorana mass arises
and are independent of
$m_{L_4}$. In the standard model
with only the fourth generation fermions and
the $4^{th}$ generation right-handed neutrino $\nu_{4 R}$,
the $\nu_4$ has the following mass matrix (ignoring mixing with lower
generation neutrinos)
$${
\pmatrix{ 0 & m_4 \cr
         m_4 & M_4 \cr},
}\eqno(2)$$
\no where $m_4$ denotes the Dirac mass induced
by electroweak breaking and $M_4$ is the Majorana mass which is allowed in
the standard model.
The eigenvalues are
$${
m_{\pm} = {1\over 2} ( M_4 \pm {\sqrt { M_4^2 + 4 m_4^2 } } )
,}\eqno(3)$$
\no and the eigenstate are charactrized by the mixing
angle
$${
\tan 2\theta = {2m_4 \over M_4}.
}\eqno(4)$$
\no Notice that the LEP Z-width constraints implies
that $m_{\pm}$ should be heavier than about $M_Z /2$.

Let us now entertain the possibility that the Majorana mass of $\nu_{4 R}$
arises from spontaneous breaking of either global or local
B-L symmetry. In what follows, we will show that the first possibility is
disfavored by astrophysical considerations.
 In the case of the second
 possibility, barring unnatural fine tuning of parameters, the scale of
local B-L must have an upper-bound, if the Yukawa couplings are assumed
to remain perturbative.

\no {\bf 2.} First we consider the case of spontaneously broken global B-L
symmetry. In this case, there will emerge the massless particle, the
Majoron[3].
We will show that
for reasonable choice
of parameters the coupling of the Majoron to electron, up and down quarks
in the presence of fourth generation neutrino is so large that it is in
 conflict with the bounds implied by the observed red giant aboundances in the
universe. This result is independent of whether the breaking of global B-L
symmetry is dynamically induced or realized by an explicit scalar field.
We will demonstrate our result using the latter example. The
model is a simple extension of the CMP model[3]
with the fourth generation fermions included and has been discussed in recent
literature[4]. To remind the reader, the CMP model is the simplest extension
of the standard model that adds only one complex, lepton-number carrying
iso-singlet scalar field $\Delta$
and one right-handed neutrino per generation.
 The field $\Delta$ couples only to the right-
handed neutrinos and when it acquires a non-zero vev ${\it {i.e.}} ~
< \Delta > = {1\over \sqrt{2} } v_{BL}$, it breaks the global B-L
symmetry spontaneously and gives rise to the massless Majoron. In the
polar decomposition of $\Delta$ {\it i.e.}
$${
\Delta = {1 \over {\sqrt 2} } ( \rho + v_{BL} ) e^{ i {\chi \over v_{BL}}}
, }\eqno(5)$$
\no $\chi$ corresponds to the Majoron.
In order to study the properties of the Majoron, let us write down the leptonic
part of the lagrangian involving only the $4^{th}$ generation fermions:
$${\eqalign{
{\cal L}= & - {\overline{L_4}} \gamma_\mu D^\mu L_4 - {\overline{\nu_{4 R}}}
                     \gamma_\mu {\partial}^\mu \nu_{4 R} \cr
          & + h_4 {\overline{L_4}}\phi \nu_{4 R} +
              f_4 \nu_{4 R}^T C^{-1} \nu_{4 R} \Delta + h.c.\cr
           & - \partial_\mu \Delta^* \partial^\mu \Delta - V( \phi,~ \Delta)
        \cr}.
}\eqno(6)$$

\no
where $\phi$ is the Higgs doublet of the standard model with vev $v_{wk} \simeq
250 GeV$.
 (Note that terms such as $\Delta^3 , ~\Delta^2 \Delta^*, etc$ are forbidden
from appearing in
$ V( \phi, \Delta )$
 by lepton number conservation.)
The massless $\chi$ can then be emitted from the stars via
processes like $ \gamma + e \rightarrow \chi + e, etc$, providing
the stars with an extra energy loss mechanism.
Obviously, large values of the electron and quark couplings to
$\chi$ would shorten the life of stars[5].
In the CMP model, in the tree-level, there is no Majoron coupling to
charged fermions. However, the neutrino exchange yields
at one-loop level
an effective
operator:
$${
\epsilon ~ M_Z~ \partial_\mu \chi ~ Z^\mu ~, }
\eqno(7)$$
\no which lead to  ${\overline e}
\gamma_5 e \chi,~
{\overline u} \gamma_5 u \chi, ~ {\overline d} \gamma_5 d \chi$ couplings.
 The upper limit on these couplings
implied by the astrophysical considerations gives[5]
$${
\epsilon~ < ~ 10^{-7} .}
\eqno(8)$$
We have performed a detail calculation of the parameter $\epsilon$
at the one-loop level
in the presence of the fourth generation neutrino with
the mass matrix given in eq.(2).
 It is finite and
expressed as follows
$${
\eqalign{
  \epsilon = & {\sin^2 2\theta \over 8 \pi^2} {M_4 \over v_{BL} }
            \{ {m_{-}\over v_{wk}} \int^1_0 dx (1-x) \ln
            {m_{-}^2 -x m_{-}^2  + x m_{+}^2 \over m_{-}^2 }\cr
            & +{ m_{+}\over v_{wk}} \int^1_0 dx ~x \ln
             { m_{-}^2 -x m_{-}^2  + x m_{+}^2 \over m_{+}^2 } \} ,\cr}
}\eqno(9)$$
\no where $m_{\pm}$
and
$\theta$ are defined in eq.(3) and (4) respectively.
\no In the limit $M_4 \gg m_4$ as usually assumed in the see-saw mechanism,
$\epsilon$ is reduced to

$${
\epsilon ~ = ~ {1\over {8 \pi^2 }} {m_4^2 \over {v_{wk} v_{BL}}}
, }\eqno(10)$$

\no This is the result obtained in ref.[3].
For the first three generations, the largest value of $m_4$ is likely to be
of the order of 1-2 GeV, in which case
the astrophysical bound is satisfied for
$v_{BL} \simeq O (TeV)$.
Let us consider the constraints on the properties of the fourth
generation neurtino resulting from LEP Z-width measurement and
the constraint in eq.(8). Firstly we know that
the scale of B-L symmetry breaking cannot be much bigger than the electroweak
scale for $f_4 \sim 1$, otherwise the constraints from LEP Z-width
measurement will not be satisfied. However, this is in contradiction
with the requirement from eq.(8), which needs a large $v_{BL}$.
 So our
arguments make highly unlikely the possibility that lepton
number breaking associated with
the fourth generation is realized via the Nambu-Goldstone
mode.

 If on the other hand, the
B-L symmetry is local, then the astrophysical constraint does not apply.
The Z-width constraint implies that the scale of local B-L
symmetry breaking should be less than a few TeV,
otherwise the see-saw mechanism would lead to a light neutrino with
mass less than $M_Z /2$.
 It therefore appears
that the existence of a fourth generation lepton with Majorana
neutrino highly constrain the scale of B-L symmetry breaking.

 The only way
to avoid these constraints would be to have an unnaturally
small value for the coupling $f_4$, in which case both constraints in the
global symmetry case could be satisfied.
For instance, if $f_4 \leq 10^{-5}$, then
a $v_{BL} \geq 10^{7}$ GeV would satisfy both constraints.

 Let us now see the constraint on $M_4$, when $m_4 \gg
M_4$. In this case,
$\epsilon$ is reduced to
$${
\epsilon = {1\over 12 \pi^2} {M_4^2 \over v_{BL} v_{wk} }
. }\eqno(11)$$
\no Taking $v_{BL} \sim 1 TeV$ as suggested above
 by the astrophysical constraint
for the first three generations, one has, from eq.(8)
$${
M_4 < O(1GeV).
}\eqno(12)$$
\no Thus the fourth generation neutrino, if it exists, will be
a Dirac or a Pseudo-Dirac
neutrino.

Our result has direct bearing on a recently proposed
dynamical symmetry breaking model
 by Hill, Luty and Paschos[6].
In their model
global B-L symmetry
is dynamically broken
near electroweak scale by neutrino (fourth generation's) condensate.
As we argued above,
 the astrophysical bounds are
not satisfied in this model. We should also point out that
these considerations can be extended to composite models and other
models with the dynamical broken symmetries, where ultra-light
Pseudo-Goldstone bosons will mix with Z directly and/or indirectly.

\no {\bf 3.}
In summary, we have examined
LEP Z-width measurement and
astrophysical constraints on the
properties of the heavy majorana neutrino, which
get mass from spontaneously B-L breakdown. We have argued that if the
$4^{th}$ generation exists, it cannot be naturally embedded
into the singlet majoron model.
In the case where B-L is a local symmetry, its breaking
scale must be less than a few TeV.
\bb
\filbreak

\ce{\bf Acknowledgements}
X.Z. thanks K.S. Babu, G. Gat and J.C. Pati for useful discussions.

\bb
\b

\ce{\bf References}
\b
\item{[1]}L. Rolandi, {\it {Talk given at the XXVI ICHEP 1992, Dallas, USA}}
     CERN-PPE/92-175, 15 October 1992.

\item{[2]}K. Olive et al, Phys. Lett. $\bf 236B$, 454 (1990);
T. Walker, G. Steigman, D. Schramm, K. Olive and H. Kang, APJ, 376, 51 (1991);
G. Steigman, The Ohio State University Preprint,
OSU-TA-11-92 (1992).

\item{[3]} Y. Chikashige, R.N. Mohapatra and R.D. Peccei,
Phys. Lett. $\bf 98B$, 265 (1980).

\item{[4]}
       C. Hill, M. Luty and E. Paschos,
Phys. Rev. $\bf D43$, 3011 (1991); C. Hill and E. Paschos, Phys. Lett.
$\bf 241B$, 96 (1990); W. Hong, D. Cline and Y. Jung, UCLA-APH 0047-5/92.

\item{[5]}D. Dicus, E. Kolb, V. Teplitz and R. Wagoner,
   Phys. Rev. $\bf D22$, 839 (1980); M. Fukugita, S. Watamura and
     M. Yoshimura, Phys. Rev. $\bf D26$, 1840 (1982);
   D. Dearborn, D. Schramm and G. Steigman, Phys. Rev. Lett,
$\bf 56$, 26 (1986);
G.G. Raffelt, Phys. Rep. $\bf 198$, 1 (1990).

\item{[6]}
C. Hill, M. Luty and E. Paschos, in Ref.[4].

\bye